\documentclass[aps,superscriptaddress,groupedaddress,twocolumn,floatfix,10pt]{revtex4}
\usepackage{natbib}
\usepackage{graphicx} 
\usepackage{dcolumn} 
\usepackage{bm} 
\usepackage{amssymb,amsmath} 
\usepackage{textcomp} 
\usepackage[utf8x]{inputenc}
\usepackage{braket}
\usepackage{amsmath}
\usepackage{placeins}
\usepackage{xcolor}
\usepackage{pdfpages}
\usepackage{soul}
\usepackage{hyperref}       
\usepackage{url}            
\usepackage{booktabs}       
\usepackage{todonotes} 
\usepackage{changebar}
\usepackage{changes}
\definechangesauthor{AB}
\begin{document}

\pagestyle{empty}

\title{Mesoscale impact of trader psychology on stock markets: a multi-agent AI approach}

\author{
Johann Lussange\\
Laboratoire des Neurosciences Cognitives, INSERM U960, Département des Études Cognitives\\
École Normale Supérieure, 29 rue d'Ulm, 75005, Paris, France.\\
Stefano Palminteri\\
Laboratoire des Neurosciences Cognitives, INSERM U960, Département des Études Cognitives\\
École Normale Supérieure, 29 rue d'Ulm, 75005, Paris, France.\\
Sacha Bourgeois-Gironde\\
Institut Jean-Nicod, UMR 8129, Département des Études Cognitives,\\
École Normale Supérieure, 29 rue d'Ulm, 75005, Paris, France.\\ 
Laboratoire d'Économie Mathématique et de Microéconomie Appliquée, EA 4442\\
Université Paris II Panthéon-Assas, 4 rue Blaise Desgoffe, 75006, Paris, France.\\
Boris Gutkin\\
Laboratoire des Neurosciences Cognitives, INSERM U960, Département des Études Cognitives\\
École Normale Supérieure, 29 rue d'Ulm, 75005, Paris, France.\\ 
Center for Cognition and Decision Making, Department of Psychology\\
NU University Higher School of Economics, 8 Myasnitskaya st., 101000, Moscow, Russia.\\
}

\date{\today}

\begin{abstract} 
Recent advances in the fields of machine learning and neurofinance have yielded new exciting research perspectives in practical inference of behavioural economy in financial markets and microstructure study. We here present the latest results from a recently published stock market simulator built around a multi-agent system architecture, in which each agent is an autonomous investor trading stocks by reinforcement learning (RL) via a centralised double-auction limit order book. The RL framework allows for the implementation of specific behavioural and cognitive traits known to trader psychology, and thus to study the impact of these traits on the whole stock market at the mesoscale. More precisely, we narrowed our agent design to three such psychological biases known to have a direct correspondence with RL theory, namely delay discounting, greed, and fear. We compared ensuing simulated data to real stock market data over the past decade or so, and find that market stability benefits from larger populations of agents prone to delay discounting and most astonishingly, to greed. 
\end{abstract}

\maketitle


\section{Introduction}

\textit{Past challenges}: Understanding how markets behave has been one of the central questions in financial market economics. Traditionally, market dynamics are studied as phenomena in themselves with a top-down approach to complexity inference, for example by using statistical or econometric models~\cite{Greene2017}. Yet trading in real financial markets comes as a result of the collective interactions of human actors, either directly in the form of economic traders, or indirectly in the form of investors’ imperatives that constrain algorithmic trading strategies~\cite{Boero2015}. The gap between these two approaches can potentially be bridged using a new generation of multi-agent systems (MAS), or agent-based models (ABM), which have been sought after by industry practitioners and regulators alike~\footnote{In the words of J.-C. Trichet, President of the European Central Bank during the 2008 financial crisis: “\textit{As a policy-maker during the crisis, I found the available models of limited help. In fact, I would go further: in the face of the crisis, we felt abandoned by conventional tools.} [...] \textit{Agent-based modelling dispenses with the optimisation assumption and allows for more complex interactions between agents.}”}. Modern MAS applied to financial markets have been able to reenact the so-called market \textit{stylised facts}~\cite{Barde2015}, which are basic aspects of the market microstructure. Neurofinance studies examined cognitive biases in individual financial decision making~\cite{Eickhoff2018}. Yet no study to our knowledge has revealed the global impact of individual cognitive traits/biases in large economic agent populations on the quantifiable financial market dynamics~\cite{Demartino2013}. 

\textit{New prospects}: The pertinence of such studies is increased by the fact that the fundamental and structural causes to the 2008 financial crisis have not been fully eliminated, and are just as potent and impactful to financial markets as they were a decade ago~\footnote{Recently, Mr. Trichet was quoted as saying “\textit{We are today in a worse situation than in 2008},” Ecorama on 11/03/2019.}. Yet, recent trends give ABM research in economics a whole new potential range of realism, coming from the association of two present-day major scientific breakthroughs: i- the steady advances of cognitive neuroscience and neuroeconomics~\cite{Eickhoff2018,Konovalov2016}, and ii- the progress of machine learning due to the increasing computational power and use of big data methods~\cite{Silver2018}. Even more promising is the synergy of these two fields, with the emergence of machine learning algorithms incorporating decision-theoretic features from neuroeconomics~\cite{Lefebvre2017,Palminteri2015}, or neuroscience models approached from the angle of machine learning~\cite{Duncan2018,Momennejad2017}. We have designed such a MAS stock market simulator~\cite{Lussange2019}, where the agents are endowed with realistic learning dynamics (each agent perform reinforcement learning in order to forecast stock prices and manage its portfolio) and identified cognitive biases and social interactions. We here extend such a platform to investigate what specific roles the cognitive biases play in emergent collective market dynamics.  

\textit{Methodology}: Such a MAS simulator emulates the microstructure of a financial stock market through a bottom-up approach to system complexity, via economic agents (e.g. investors, institutions) and their economic transactions (e.g. buying, selling, holding stocks). Each individual agent is modelled according to three distinct features: i- a reinforcement learning algorithm to develop its own skills for price forecasting and stock trading (i.e. each agent learns to trade over time), ii- such an agent learning process framed for a rather chartist or fundamentalist approach to stock price valuation, iii- implementation of behavioural and cognitive biases relevant to trader psychology, as the framework of reinforcement learning is known to have multiple parallels with decision processes in the brain~\cite{Momennejad2017}, and computationally offers the possibility to functionally embed decision theory features to model specific neuroeconomic biases. At each time step of the simulation, the agents individually learn to either buy, sell or hold stocks, in a given number and at a given price. To do this, they each send the transaction orders to an order book that is common to all agents and which sorts and matches the transaction orders received. The agents learn the outcome (or reward) of their economic investments over time. The law of supply and demand, and other key phenomena to price formation~\cite{Dodonova2018,Naik2018} such as illiquidity and bid-ask spread are thus reenacted. The simulated price and volume time series of each stock are compared to real financial data. The MAS parameters can be calibrated so that its output matches real stock markets, thereby measuring the collective role played by the cognitive traits of the agents in a quantitative way, as described in Section IV.

\section{Reinforcement learning}

\textit{Overview}: We here briefly review the basics of RL theory that pertain to this study, and how these have a rich correspondence with neuroscience research. Together with supervised and unsupervised learning, RL has been termed one of the three paradigm shifts of machine learning, and is today at the forefront of almost all breakthroughs in AI research. Like many other machine learning methods, RL has its roots in behavioural psychology and decision theory. In RL, we consider an agent in a given environment that must learn the best way to consistently receive a preset reward from its environment through its possible actions. The whole RL problem and solution is thus how the agent matches these actions in a dynamic environment so as to maximise this reward. In the beginning of the task, the agent is completely agnostic as to which actions are best to use: it will have to learn this on its own. 

\vspace{1mm}

\textit{Parameters}: The RL problem is defined with three main parameters: the states of the environment $s \in \mathcal S$, the agents actions $a \in \mathcal A$, and the agent reward $r \in \mathbb R$. The states and actions can be defined as more or less complex concepts, and notice the rewards can be positive or negative. The goal of RL for the agent is to find its policy $\pi(s,a) = Pr (a | s)$ so as to maximise its rewards. In order to do this, two major types of RL algorithms are used: i- model-based methods rely on the agent estimating two functions called the transition probability $\mathcal P_{ss'}^a = Pr \{ s_{t+1} = s' | s_t=s, a_t=a \}$ and the expected value $\mathcal R_{ss'}^a = \mathbb E [ r_{t+1} | s_t=s, a_t=a, s_{t+1}=s' ]$, where $0 < \gamma < 1$ is a discount parameter related to the concept of delayed reward, and out of these derive the so-called \textit{state-value function}: $V(s)= \mathbb E [ \sum_{k=0}^{\infty} \gamma^k r_{t+k+1} | s_t=s ]$. ii- Model-free methods rely more simply on the estimation of the so-called \textit{action-value function} $Q(s,a)= \mathbb E [ \sum_{k=0}^{\infty} \gamma^k r_{t+k+1} | s_t=s, a_t=a ]$. These functions $V(s)$ and $Q(s,a)$ in model-based and model-free methods thus allow the agent to update its policy, which in turn shall be used at the next time step of the task to select a relevant action $a$, and iteratively proceed in a same manner. 

\vspace{1mm}

\textit{Features}: Three major features appear here, and are at the centre of all RL research: i- \textit{curse of dimensionality} arises from the number of state-action pairs. If these are two numerous, the problem of convergence to a policy may be intractable. ii- \textit{Temporal credit assignment} is another issue pertaining to how rewards are practically defined for the task at hand, and how temporal discounting of these rewards is set. iii- \textit{Exploration vs. exploitation} is the last issue, which pertains to whether it is profitable to the agent to exploit the rewards linked to a good policy it found in its environment, or whether it is better to continue exploring and (perhaps) attain to a better policy and hence rewards.

\section{Model}

\textit{Features}: We have designed such a MAS stock market simulator, and published the main aspects of its performance and calibration process to real stock market data in a previous work~\cite{Lussange2019}, to which we refer the reader for more details, as we will simply here recall the general frame of its architecture. This MAS simulator comprises three novel features: i- each agent learns to forecast and trade autonomously by reinforcement learning with a long-only equity strategy, ii- the agent discretionary asset pricing process relies on learning to weight both chartist and fundamentalist inputs, iii- the agents reinforcement learning framework is embedded with specific cognitive and behavioural traits proper to trader psychology, which for the sake of simplicity and robustness we constrained to three: delay discounting, fear, and greed. The latter features allow us to study the weight and impact of certain agent psychological traits on stock markets at the mesoscale, by comparing our simulated data with real stock market data, coming from the London Stock Exchange over the years $2007$ to $2018$. 

\vspace{1mm}

\textit{Architecture}: Let's first briefly outline the general profile of this MAS stock market simulator, which generally deals with a set of two C++ classes: i- a number $I$ of agents which try and maximise over time the net asset value of their individual portfolio, which consists in risk-free assets (bonds) and a number of stocks (equity), and ii- a number $J$ of different double-auction limit order books, each corresponding to the transactions of a stock $j \in J$ at each time step of the simulation. At the beginning of the simulation, each agent is completely agnostic wrt. both price forecasting and trading. It will autonomously learn these by two distinct reinforcement learning algorithms, and send at at each time step $t$ of the simulation a transaction order to the order book concerning a specific number of each stock $j$ to buy or sell, or will simply hold its position and wait for a better time to trade. At each time step, each order book thus collects the transaction orders of all agents and processes them by sorting the bid orders in a descending way, and the ask orders in an ascending way, matching them for transactions at mid-price at each level, starting from the top, until bids no longer exceed offers. That latest transaction at the lowest possible level is then the market price of stock $j$ at next time step $P^{j}(t+1)$, the total number of stocks transacted is its traded volume at next time step $V^{j}(t+1)$, and the absolute difference between the average of all bids and asks is its spread $S^{j}(t+1)$. The market price at time $t=0$ is set by default at $P(t=0)=\pounds 100$, but for their own asset pricing, agents approximate by cointegration~\citep{Murray1994} another time series $\mathcal{T}^{j}(t)$ generated at time $t=0$, which corresponds to the fundamental prices of asset $j$, as in other models~\cite{Franke2011,Chiarella2007}. The inspiration for the profile of such data is a metric often encountered in corporate finance called \textit{enterprise value}~\citep{Vernimmen2017}, which is the theoretical price at which the company issuing the stocks would be acquired, and which can give a rough fundamental stock price estimate, if divided by the total number of stocks outstanding. The weight given to this fundamentalist valuation or to the market price is learned by each agent, so that some agents will tend to be more chartists or fundamentalists. Agents trade as such for a learning phase of $1000$ time steps, after which all their portfolio assets are reset to their initial values, and the simulation then let to run for statistical inference and microstructure study. Past this learning phase, we consider a simulation of $T$ time steps, where one time step typically represents a trading day, and thus $T_w=5$, $T_m=21$, $T_y=286$ correspond to a trading week, month, and year, respectively. 

\vspace{1mm}

\textit{Agents}: We here give more details on the agents design, which are initialised with specific parameters. These execute at each time step $t$, and for each stock $j$, a reinforcement learning algorithm $\mathcal{F}^{i}$ to do price forecasting, and another one to learn trading $\mathcal{T}^{i}$ based on the result of $\mathcal{F}^{i}$. 

i- \textit{Parameters}: With its number of initial stocks, cash reserves, and proper psychological traits of behaviour and cognition (see below), each agent is initialised at time $t=0$ with specific parameters such as: an investment horizon $\tau^{i} \sim \mathcal{U} \{T_w, 6T_m \}$ (corresponding to the number of time steps after which the agent liquidates its position), a memory interval $h^{i} \sim \mathcal{U} \{ T_w, T\}$ (corresponding to the size of the rolling time interval used by the agent for its learning process), a transaction gesture $g^{i} \sim \mathcal{U} (0.2, 0.8)$ (scaling with the spread and related to how far above or below the value of its own stock pricing the agent is willing to trade and deal a transaction), or a reflexivity amplitude parameter $\rho^{i} \sim \mathcal{U} (0, 100\%)$ (gauging the weight given by the agent to fundamental or chartist valuation of the stock). 

ii- \textit{Forecasting}: The reinforcement learning states of the forecasting algorithm $\mathcal{F}^{i}$ are: a longer-term price volatility $s_0^{\mathcal{F}}$ ($0$ for low, $1$ for mid, $2$ for high), a shorter-term price volatility $s_1^{\mathcal{F}}$ ($0$ for low, $1$ for mid, $2$ for high), and the gap between its own present fundamental valuation and the present market price $s_2^{\mathcal{F}}$ ($0$ for low, $1$ for mid, $2$ for high). Out of these states, the agent chooses an action in order to optimise its price prediction at its investment horizon $\tau^{i}$: the type of econometric forecast $a_0^{\mathcal{F}}$ ($0$ for mean-reverting, $1$ for averaging, $2$ for trend-following), the size of the historical lag interval for this econometric forecast $a_1^{\mathcal{F}}$ ($0$ for short, $1$ for mid, $2$ for large), and the weight given to its reflexivity amplitude parameter $\rho^{i}$ for price estimation $a_2^{\mathcal{F}}$ ($0$ for low, $1$ for mid, $2$ for large). The rewards of these performed actions are defined via the mismatches between past forecasts at time $t-\tau^{i}$ and their eventual price realisation at time $t$. These feed a direct policy update with new action probabilities for the agent in such a state. 

iii- \textit{Trading}: The reinforcement learning states of the forecasting algorithm $\mathcal{T}^{i}$ are: the trend of the price forecast of the previous algorithm $s_0^{\mathcal{T}}$ ($0$ for decreasing, $1$ for stable, $2$ for increasing), the price volatility $s_1^{\mathcal{T}}$ ($0$ for low, $1$ for mid, $2$ for high), the level of the agent risk-free assets compared to its initial values $s_2^{\mathcal{T}}$ ($0$ for low, $1$ for high), the level of the agent stock holdings compared to its initial values $s_3^{\mathcal{T}}$ ($0$ for low, $1$ for high), and the stock liquidity based on previous exchanged volumes $s_4^{\mathcal{T}}$ ($0$ for zero, $1$ for low, $2$ for high). From this state, the agent can chose the following actions: sending an order to the order book $a_0^{\mathcal{T}}$ ($0$ for shorting, $1$ for holding, $2$ for longing), and at what price above or below the agent's own price estimate $a_1^{\mathcal{T}}$ ($0$ for indifference to lose on transaction, $1$ for neutral, $2$ for willingness to gain on transaction) via the transaction gesture $g^{i}$ scaled with the market spread $S^{j}(t)$. The rewards of these performed actions are defined via the difference in cashflow at time $t$ between the profit or loss consequent to the agent's past action at time $t-\tau^{i}$, and the one had this action not been taken. Again, these feed a direct policy update with new action probabilities for the agent in such a state.

\section{Agent psychology}

\textit{Considerations}: We now want to gauge the impact of trader psychology on stock markets. Today, the proportion of trading in financial stock markets that is of algorithmic origin is above $80 \%$ in terms of transaction cashflows on major exchanges~\cite{Mukerji2019}. One could hence rightly think about the relevance of such a study, at a time when these are increasingly subject to algorithmic trading and automated portfolio management (passive asset management, exchange-traded funds, robo-advisors, etc.). Nevertheless, the large impact played by behavioural economics and trader psychology on stock markets still remains, for the following reason: the performance constraints (risk, return, liquidity, etc.) placed on these algorithms are (and will always be) of human origin. For example, the relevance of time-discounting is seen in practical deadlines assigned by investors to portfolio managers to assess asset management performance. Or we can say the same about fear and greed via the use of common thresholds for Sharpe ratio metrics~\cite{Bayley2012}, and the choice of investors to be more or less risk-averse for certain expected returns. Such indirect impact of human cognition and behaviour on increasingly automated stock markets is even seen in the ever-larger role played by entire classes of trading strategies, like high-frequency trading, to which portfolio managers have progressively turned because of the preference of many investors for smoother equity curves and minimal drawdowns. 

\vspace{1mm}

\textit{Traits}: The framework of RL has a direct correspondence with decision theory, and hence several traits of learning, cognition, and behaviour can be easily implemented through it~\cite{Lefebvre2017,Palminteri2015}. For example, \textit{belief revision} (which is defined as changing one's belief insufficiently in the face of new evidence) is straightforwardly modelled by the reinforcement learning rate. But in this Paper, we have for the sake of simplicity and robustness restricted our attention to three specific psychological traits, which allow us to model certain interesting and relevant issues proper to behavioural economic and investor behaviour in financial markets: 

i- \textit{Delay discounting}: This is defined as having greater economic utility on shorter time-scales than longer ones~\cite{Yanoff2015}. It is modelled by the agent being initialised with a much lower investment horizon $\tau^{i}$, now drawn from a discrete uniform distribution $\mathcal{U} [T_w, 2T_w($. Note that delay discounting may be regarded as a trait of addictive behaviour. 

ii- \textit{Fear}: This is modelled by the second reinforcement algorithm $\mathcal{T}^{i}$ taking the action $a_0^{\mathcal{T}}=0$ (to send a sell order to the order book) if either $s_1^{\mathcal{T}}=2$ (high volatility), $s_2^{\mathcal{T}}=0$ (low risk-free assets), or $s_4^{\mathcal{T}}=0$ (illiquid stock), regardless of the price direction shown by $s_0^{\mathcal{T}}$. This is performed once every $n=5$ time steps on average, according to a uniform distribution. Notice this is congruent with the long-only equity strategy of all agents. 

iii- \textit{Greed}: This is modelled by the second reinforcement algorithm $\mathcal{T}^{i}$ taking the action $a_0^{\mathcal{T}}=2$ (to send a buy order to the order book) only if $s_0^{\mathcal{T}}=2$ (trend of the price is to increase according to the forecast $\mathcal{F}^{i}$) regardless of other state indicators such as volatility or illiquidity. This is done once every $n=5$ time steps on average, according to a uniform distribution. Notice this is also congruent with the long-only equity strategy of all agents.  

These traits can be implemented in a chosen percentage of the agents population, and one can observe and study the ensuing changes in mesoscale impact on the market by varying these percentages. As in our previous work~\cite{Lussange2019}, we compare these results to real stock market data, which comes from $640$ stocks, that have been continuously traded on the London Stock Exchange over the years $2007$ to $2018$.

\section{Results}

\subsection{Delay discounting}

\begin{figure}[!htbp]
\includegraphics[scale=0.53]{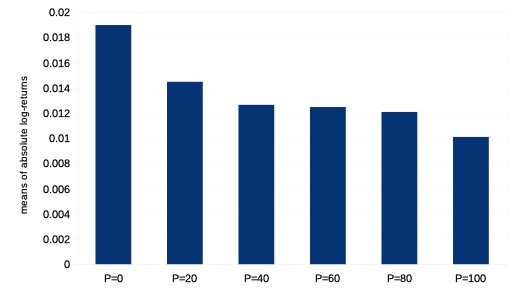}
\caption{\label{B1} Means of all absolute logarithmic price returns, for real data, and simulations with a percentage $p$ of agents corresponding to $p=0\%, 20\%, 40\%, 60\%, 80\%, 100\%$ of the total agent population with a delay discounting profile (the remainder $100-p$ being bias-free agents). The simulations are generated with parameters $I=500$, $J=1$, $T=2875$ (corresponding to about $11$ years), and $S=20$.}
\end{figure}

\begin{figure}[!htbp]
\includegraphics[scale=0.53]{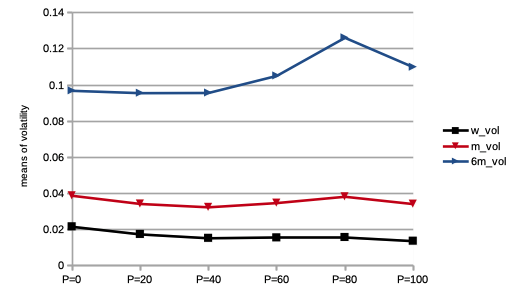}
\caption{\label{B2} Means of all volatilities (defined as standard deviations of price normalised to price itself $\sigma/P(t)$) computed over lags of one weeks (black), one month (red), and six months (blue) intervals, for simulations with a percentage $p$ of agents corresponding to $p=0\%, 20\%, 40\%, 60\%, 80\%, 100\%$ of the total agent population with a delay discounting profile (the remainder $100-p$ being bias-free agents). The simulations are generated with parameters $I=500$, $J=1$, $T=2875$ (corresponding to about $11$ years), and $S=20$.}
\end{figure}

\begin{figure}[!htbp]
\includegraphics[scale=0.53]{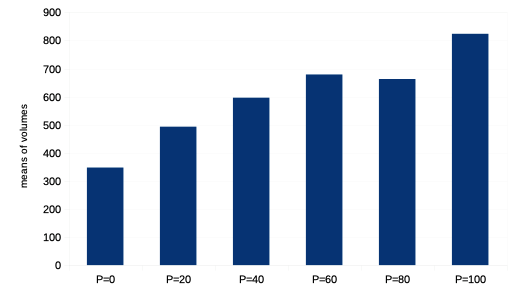}
\caption{\label{B3} Means of all trading volumes, for a percentage $p$ of agents corresponding to $p=0\%, 20\%, 40\%, 60\%, 80\%, 100\%$ of the total agent population with a delay discounting profile (the remainder $100-p$ being bias-free agents). The simulations are generated with parameters $I=500$, $J=1$, $T=2875$ (corresponding to about $11$ years), and $S=20$.}
\end{figure}

\begin{figure}[!htbp]
\includegraphics[scale=0.53]{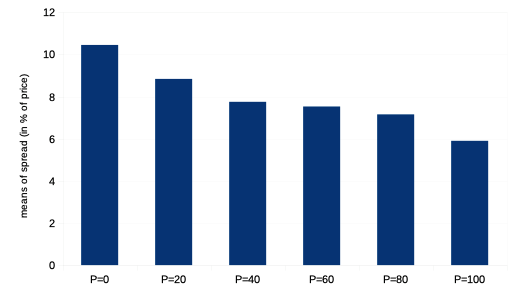}
\caption{\label{B4} Means of all bid-ask spread in percent of price, for a percentage $p$ of agents corresponding to $p=0\%, 20\%, 40\%, 60\%, 80\%, 100\%$ of the total agent population with a delay discounting profile (the remainder $100-p$ being bias-free agents). The simulations are generated with parameters $I=500$, $J=1$, $T=2875$ (corresponding to about $11$ years), and $S=20$.}
\end{figure}

\begin{figure}[!htbp]
\includegraphics[scale=0.53]{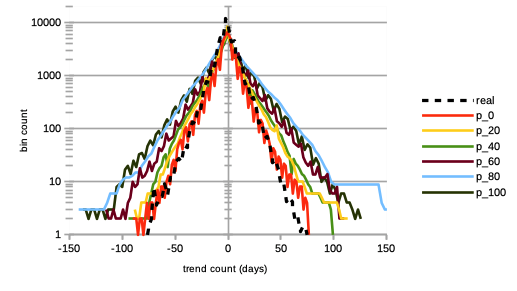}
\caption{\label{B5} Distribution of the number of consecutive days of rising prices (positive values) and dropping prices (negative values). This is for both real (dashed black curve) and simulated (continuous curves) data, the latter being for a percentage $p$ of agents corresponding to $p=0\%$ (red), $p=20\%$ (yellow), $p=40\%$ (green), $p=60\%$ (brown), $p=80\%$ (light blue), and $p=100\%$ (dark green) of the total agent population with a delay discounting profile (the remainder $100-p$ being bias-free agents). The simulations are generated with parameters $I=500$, $J=1$, $T=2875$ (corresponding to about $11$ years), and $S=20$.}
\end{figure}

\begin{figure}[!htbp]
\includegraphics[scale=0.53]{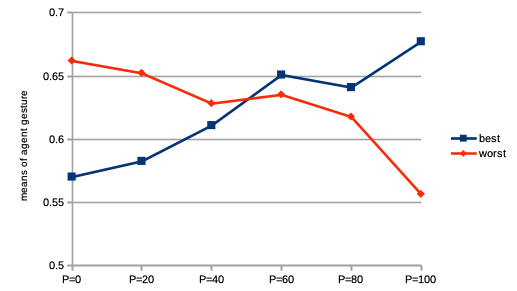}
\caption{\label{B6} Means of individual agents gesture $g^{i}$ among $10\%$ best (blue) and $10\%$ worst (red) performing agents at time $t=T$, for simulations with a percentage $p$ of agents corresponding to $p=0\%, 20\%, 40\%, 60\%, 80\%, 100\%$ of the total agent population with a delay discounting profile (the remainder $100-p$ being bias-free agents). The simulations are generated with parameters $I=500$, $J=1$, $T=2875$ (corresponding to about $11$ years), and $S=20$.}
\end{figure}

We first want to study the market impact of larger percentages of agents set with delay discounting. The interest of such a study can be linked with the role played by higher frequency trading in present stock exchanges: for when increasingly more agents trade more frequently or at shorter time scales, one could ask for instance if this is beneficiary or not to market stability. For simulations with increasing percentages of agents having such a delay discounting profile as described in the previous section, we observe the following:
\begin{itemize}
\item[--] We see on Fig. \ref{B1} a strong decrease in absolute logarithmic price returns. 
\item[--] We see on Fig. \ref{B2} a steady decrease in short-term price volatility, and increase in long-term volatility. 
\item[--] We see on Fig. \ref{B3} a very strong increase in trading volumes, as expected. 
\item[--] We see on Fig. \ref{B4} a steady decrease in market bid-ask spread. 
\item[--] We see on Fig. \ref{B5} a greater propensity for longer bull and bear market regimes. 
\item[--] We see on Fig. \ref{B6} a reversion effect by which greater propensity for best performing agents to have a large transaction gesture $g^{i}$, and conversely for worst performing agents to have a smaller transaction gesture. 
\item[--] The rates of agent bankruptcy remain stable regardless of these varying percentages. 
\end{itemize}

Increasing numbers of such delay discounting agents are thus posited to be beneficial to stock market stability, by lowering general volatility and increasing trading volumes, and thus tackling the issue of market illiquidity propitious to crashes.

\subsection{Fear}

\begin{figure}[!htbp]
\includegraphics[scale=0.53]{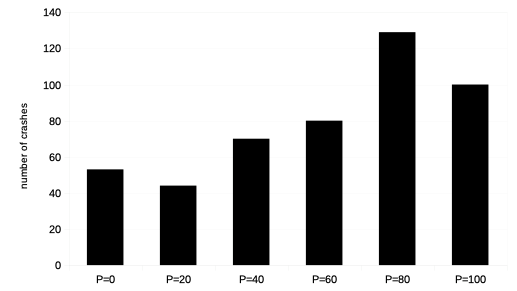}
\caption{\label{D1} Number of market crashes (defined as a drop of more than $20\%$ in market price), for a percentage $p$ of agents corresponding to $p=0\%, 20\%, 40\%, 60\%, 80\%, 100\%$ of the total agent population with a fear profile (the remainder $100-p$ being bias-free agents). The simulations are generated with parameters $I=500$, $J=1$, $T=2875$ (corresponding to about $11$ years), and $S=20$.}
\end{figure}

\begin{figure}[!htbp]
\includegraphics[scale=0.53]{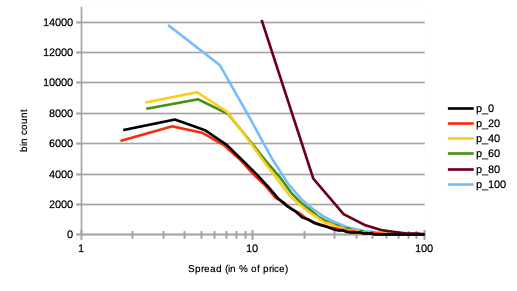}
\caption{\label{D2} Distribution of bid-ask spread in percent of price, for a percentage $p$ of agents corresponding to $p=0\%$ (black), $p=20\%$ (red), $p=40\%$ (yellow), $p=60\%$ (green), $p=80\%$ (brown), and $p=100\%$ (light blue) of the total agent population with a fear profile (the remainder $100-p$ being bias-free agents). The simulations are generated with parameters $I=500$, $J=1$, $T=2875$ (corresponding to about $11$ years), and $S=20$.}
\end{figure}

We then want to study the market impact of larger percentages of agents set with fear as a psychological trait. Fear and greed have both been described as the main forces behind stock market dynamics, and especially in specific market regimes, such as bubbles or crashes, the propensity of investors to fear or even panic has a large game theoretic impact on stock market dynamics and stability. For larger percentages of agents with such a fear profile as described in the previous section, we observe the following: 
\begin{itemize}
\item[--] We see on Fig. \ref{D1} a sharp increase in stock market crashes, as expected.
\item[--] We see on Fig. \ref{D2} a steady increase in market bid-ask spread.
\end{itemize}

We shall also add that trading volumes slowly increase, and that the proportion of bankrupt agents remains stable with larger populations of agents with such a fear profile.

\subsection{Greed}

\begin{figure}[!htbp]
\includegraphics[scale=0.53]{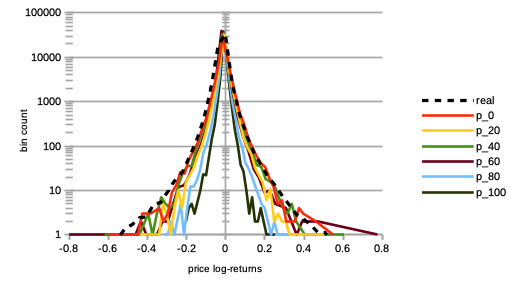}
\caption{\label{C1} Distribution of logarithmic returns of prices $\log [P(t)/P(t-1)]$ of real (dashed black curve) and simulated (continuous curves) data. The simulations are for a percentage $p$ of agents corresponding to $p=0\%$ (red), $p=20\%$ (yellow), $p=40\%$ (green), $p=60\%$ (brown), $p=80\%$ (light blue), and $p=100\%$ (dark green) of the total agent population with a greed profile (the remainder $100-p$ being bias-free agents). The simulations are generated with parameters $I=500$, $J=1$, $T=2875$ (corresponding to about $11$ years), and $S=20$.}
\end{figure}

\begin{figure}[!htbp]
\includegraphics[scale=0.53]{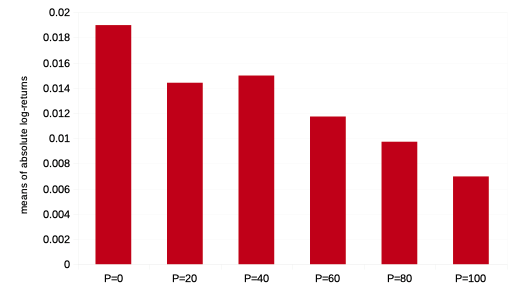}
\caption{\label{C2} Means of all absolute logarithmic price returns, for real data and simulations with a percentage $p$ of agents corresponding to $p=0\%, 20\%, 40\%, 60\%, 80\%, 100\%$ of the total agent population with a greed profile (the remainder $100-p$ being bias-free agents). The simulations are generated with parameters $I=500$, $J=1$, $T=2875$ (corresponding to about $11$ years), and $S=20$.}
\end{figure}

\begin{figure}[!htbp]
\includegraphics[scale=0.53]{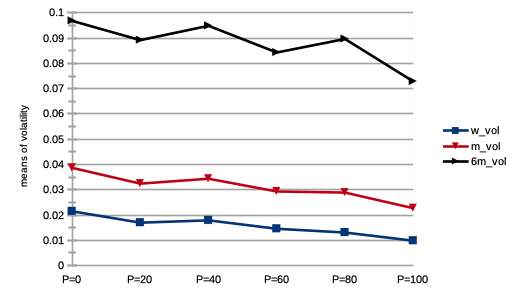}
\caption{\label{C3} Means of all volatilities (defined as standard deviations of price normalised to price itself $\sigma/P(t)$) computed over lags of one weeks (black), one month (red), and six months (blue) intervals, for simulations with a percentage $p$ of agents corresponding to $p=0\%, 20\%, 40\%, 60\%, 80\%, 100\%$ of the total agent population with a greed profile (the remainder $100-p$ being bias-free agents). The simulations are generated with parameters $I=500$, $J=1$, $T=2875$ (corresponding to about $11$ years), and $S=20$.}
\end{figure}

\begin{figure}[!htbp]
\includegraphics[scale=0.53]{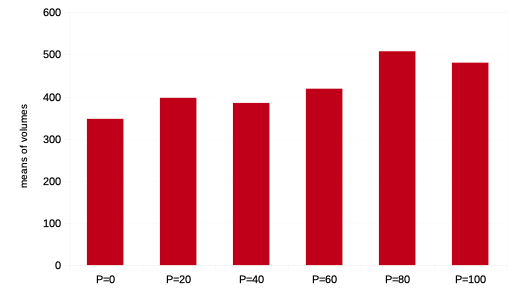}
\caption{\label{C4} Means of all trading volumes, for a percentage $p$ of agents corresponding to $p=0\%, 20\%, 40\%, 60\%, 80\%, 100\%$ of the total agent population with a greed profile (the remainder $100-p$ being bias-free agents). The simulations are generated with parameters $I=500$, $J=1$, $T=2875$ (corresponding to about $11$ years), and $S=20$.}
\end{figure}

\begin{figure}[!htbp]
\includegraphics[scale=0.53]{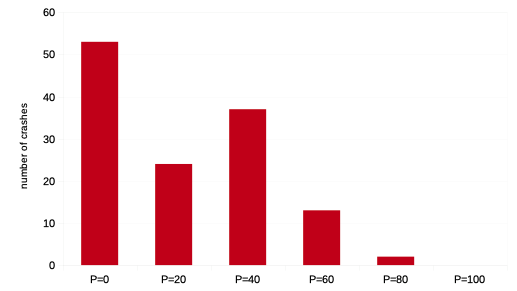}
\caption{\label{C5} Number of market crashes (defined as a drop of more than $20\%$ in market price), for a percentage $p$ of agents corresponding to $p=0\%, 20\%, 40\%, 60\%, 80\%, 100\%$ of the total agent population with a greed profile (the remainder $100-p$ being bias-free agents). The simulations are generated with parameters $I=500$, $J=1$, $T=2875$ (corresponding to about $11$ years), and $S=20$.}
\end{figure}

\begin{figure}[!htbp]
\includegraphics[scale=0.53]{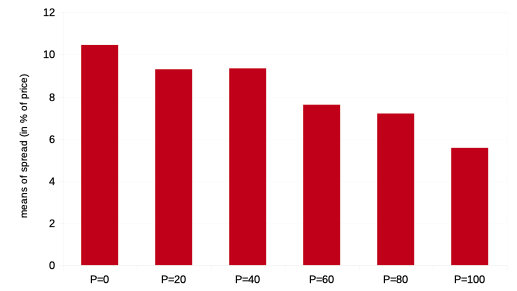}
\caption{\label{C6} Means of all bid-ask spread in percent of price, for a percentage $p$ of agents corresponding to $p=0\%, 20\%, 40\%, 60\%, 80\%, 100\%$ of the total agent population with a greed profile (the remainder $100-p$ being bias-free agents). The simulations are generated with parameters $I=500$, $J=1$, $T=2875$ (corresponding to about $11$ years), and $S=20$.}
\end{figure}

\begin{figure}[!htbp]
\includegraphics[scale=0.53]{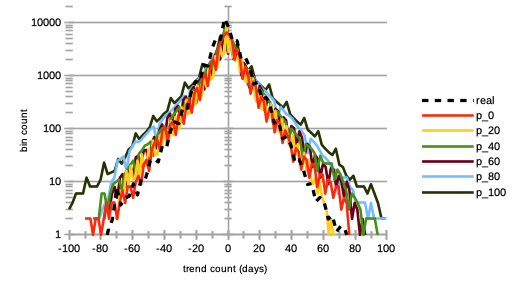}
\caption{\label{C7} Distribution of the number of consecutive days of rising prices (positive values) and dropping prices (negative values). This is for both real (dashed black curve) and simulated (continuous curves) data, the latter being for a percentage $p$ of agents corresponding to $p=0\%$ (red), $p=20\%$ (yellow), $p=40\%$ (green), $p=60\%$ (brown), $p=80\%$ (light blue), and $p=100\%$ (dark green) of the total agent population with a greed profile (the remainder $100-p$ being bias-free agents). The simulations are generated with parameters $I=500$, $J=1$, $T=2875$ (corresponding to about $11$ years), and $S=20$.}
\end{figure}

We finally want to study the impact of investor greed, as we studied fear, and assess its mesoscale influence on stock market stability. For larger percentages of agents with such a greed profile as described in the previous section, we observe the following: 
\begin{itemize}
\item[--] We see on Fig. \ref{C1} and \ref{C2} a decrease in logarithmic returns.
\item[--] We see on Fig. \ref{C3} a steady decrease in price volatilities at all time scales.
\item[--] We see on Fig. \ref{C4} a slow increase in trading volumes. 
\item[--] We see on Fig. \ref{C5} a very sharp diminution of market crashes. 
\item[--] We see on Fig. \ref{C6} a strong decrease in market bid-ask spread. 
\item[--] We see on Fig. \ref{C7} a greater likelihood for bull market regimes. 
\item[--] The rates of agent bankruptcy strongly decrease with these varying percentages, being almost halved beyond $p>80\%$.
\end{itemize}

We can thus come to the somewhat counter-intuitive conclusion that larger populations of greedy agents sharply increase market stability.

\clearpage 
\section{Conclusion}

We thus modelled a stock market via a multi-agent system, where the agents autonomously perform portfolio management via long-only equity strategies, based on autonomous reinforcement learning algorithms performing price forecasting and stock trading, with a chartist or fundamentalist approach to price estimation. In such a model, each agent is also endowed with specific and relevant psychological traits proper to behavioural economics, that can be naturally implemented via the agent reinforcement learning framework. These allow us to study the impact of agent learning on financial stock markets at the mesoscale, as we narrowed our study to three such psychological traits: delay discounting, fear, and greed. Through comparison of such simulated data, we found that market stability greatly benefits from larger numbers of such delay discounting agents. We also observe that market stability especially benefits from increasing (resp. decreasing) proportions of greedy (resp. fearful) agents. Perhaps counter-intuitively, the rates of agent bankruptcy strongly decrease with larger percentages of greedy agents, while remaining stable for larger population of agents with a fear or delay discounting profile. Future extension of this work would naturally be to study other psychological biases, but also to assess bubble regimes in stock markets. One could also extend the framework of agent trading so as to encompass other trading strategies and portfolio diversification to explore cross-asset structures. We expect these results to be of interest to financial regulatory instances as well as academia. 

\section{Acknowledgment}

We graciously acknowledge this work was supported by the RFFI grant nr. 16-51-150007 and CNRS PRC nr. 151199.

\medskip

\bibliography{Article}
 
\end{document}